\documentclass[twocolumn]{aastex63}

\usepackage{hyperref}

\usepackage{float}
\usepackage{amsmath}
\usepackage{afterpage}
\usepackage{enumitem}
\usepackage{pifont}
\renewcommand\ion[2]{#1$\;${\scshape{#2}}}
\usepackage{statmath}

\def\msun{M$_{\odot}$}
\def\lsun{L$_{\odot}$}

\def\kms{${\text{km s}^{-1}}$}

\def\farcs{\ensuremath{\overset{\prime\prime}{.}}}
\def\deg{\ensuremath{^{\circ}}}

\def\arcs{\ensuremath{^{\prime\prime}}}

\def\mjy{mJy\,beam$^{-1}$}
\def\mujy{$\mu$Jy\,beam$^{-1}$}

\begin{document}

\title{Radio Continuum and Water Maser Observations of the High-Mass Protostar IRAS 19035+0641 A}
\shorttitle{The IRAS 19035+0641 A Jet}

\author[0000-0003-0090-9137]{Tatiana M. Rodr\'iguez}
\affiliation{Physics Department, New Mexico Tech, 801 Leroy Pl., Socorro, NM 87801, USA.}
\affiliation{Student at the National Radio Astronomy Observatory, 1003 Lopezville Road, Socorro, NM 87801, USA.}

\author{Emmanuel Momjian}
\affiliation{National Radio Astronomy Observatory, P.O. Box O, 1003 Lopezville Road, Socorro, NM 87801, USA}

\author{Peter Hofner}
\affiliation{Physics Department, New Mexico Tech, 801 Leroy Pl., Socorro, NM 87801, USA.}
\affiliation{Adjunct Astronomer at the National Radio Astronomy Observatory, 1003 Lopezville Road, Socorro, NM 87801, USA.}

\author{Anuj P. Sarma}
\affiliation{Physics and Astrophysics Department, DePaul University, 2219 N. Kenmore Ave., Chicago, IL 60614, USA}

\author[0000-0001-6755-9106]{Esteban D. Araya}
\affiliation{Physics Department, Western Illinois University, 1 University Circle, Macomb, IL 61455, USA.}
\affiliation{Physics Department, New Mexico Tech, 801 Leroy Pl., Socorro, NM 87801, USA.}

\begin{abstract}
We present Very Large Array (VLA) 1.3\,cm continuum and 22.2~GHz H$_2$O maser observations of the high-mass protostellar object IRAS~19035+0641~A. Our observations unveil an elongated bipolar 1.3~cm continuum structure at scales $\lesssim500\,$au which, together with a rising in-band spectral index, strongly suggests that the radio emission toward IRAS~19035+0641~A arises from an ionized jet. In addition, eight individual water maser spots well aligned with the jet axis were identified. 
The Stokes V spectrum of the brightest H$_2$O maser line ($\sim100\,$Jy) shows a possible Zeeman splitting and is well represented by the derivatives of two Gaussian components fitted to the Stokes I profile. 
The measured $B_{\mathrm{los}}$ are $123\,(\pm27)$ and $156\,(\pm8)\,$mG, translating to a pre-shock magnetic field of $\approx7\,$mG. 
Subsequent observations to confirm the Zeeman splitting showed intense variability in all the water maser spots, with the brightest maser completely disappearing. 
The observed variability in a one-year time scale could be the result of an accretion event. These findings strengthen our interpretation of IRAS~19035+0641~A as a high-mass protostar in an early accretion/outflow evolutionary phase.
\end{abstract}


\section{Introduction}
\label{sec:intro}
The typical formation time-scale of high-mass stars ($M_*>8\,M_\odot$) is much shorter than their lower mass counterparts ($\sim10^5$ versus $\sim10^7\,$yr, e.g., \citealt{Andre94,Sabatini21}), and they are still deeply embedded in their parental cloud by the time hydrogen burning in the core ignites. 
This makes it extremely challenging to observationally study the early stages of high-mass star formation. 
According to the turbulent core accretion model \citep[e.g.,][]{Mckee03}, as the young stellar object (YSO) accretes material, jets of ionized gas are ejected from very near the protostellar surface.
Even though these ionized jets play a key role in the high-mass star formation process, their nature is still poorly understood, mainly due to a lack of sub-arcsecond resolution radio continuum data. 
Recently, \cite{Rosero16,Rosero19} carried out a deep (rms noise $\sim5\,$\mujy), high angular resolution ($\theta\sim0$\farcs33) continuum survey at 1.3 and 6~cm toward 58 high-mass star forming regions utilizing the Karl G. Jansky Very Large Array (VLA). They report the detection of 70 individual radio sources, including \ion{H}{ii} regions, ionized jets, and ionized jet candidates. The latter were classified as such based on their rising spectral index (expected from ionized jets as per \citealt{Reynolds86}) and association with usual jet tracers, such as molecular outflows, mostly from single-dish data. However, they were unresolved or slightly resolved in their observations, which made it impossible to completely discard ionization by a zero age main sequence star at the center of a hyper-compact (HC) \ion{H}{ii} region as the origin of the radio continuum emission. 
Nonetheless, their statistical analysis of ionized jets and jet candidates provided significant further evidence supporting the common origin scenario for radio jets driven by YSOs of all luminosities \citep[see Figure~8 in][]{Rosero19}. 
This, in turn, supports the idea that high- and low-mass stars are formed in a similar manner, as predicted by the turbulent core accretion model.
Ionized jets from such high-mass YSOs are expected to present an elongated morphology at scales of a few hundred au from the driving protostellar object, while \ion{H}{ii} regions are more spherical.
In order to investigate the shock ionization origin of the radio continuum sources classified as jet candidates on these scales, higher angular resolution observations are essential.  
Therefore, we carried out a $3\times$ higher resolution ($\theta\sim0$\farcs1) VLA survey at 1.3~cm toward all jet candidates from \cite{Rosero16,Rosero19}.
The angular resolution of our data translates to linear resolutions between $\sim150$ and 1200\,au (depending on the distance to each source), which will enable us to clearly discern whether the jet candidates are indeed jets or \ion{H}{ii} regions.

As a byproduct of our radio continuum survey, we observed the 22.2~GHz ($6_{1,6}-5_{2,3}$) H$_2$O (water) maser emission toward the jet candidates.
Water masers are collisionally pumped in high density regions ($n\sim10^{8-10}\,$cm$^{-3}$, \citealt{Elitzur92}) and are usually associated with jets and outflows from early stage high-mass YSOs \citep[e.g.,][]{Moscadelli19,Ladey22}. The detection of water masers in close proximity to the jet candidates would therefore strongly support the jet interpretation for the radio continuum sources.
Moreover, water masers are highly compact and can be extremely bright, which makes them an ideal probe to study shock dynamics at sub-arcsecond resolution \citep[e.g.,][]{Moscadelli20}. 
In addition, the Zeeman effect detected using H$_2$O masers offers direct measurements of the line-of-sight magnetic field strength in star forming regions \citep[e.g.,][]{Sarma08}. However, because the water molecule is weakly sensitive to the Zeeman effect (the splitting factor is $0.0021\,\text{Hz }\mu\text{G}^{-1}$, versus $2.8\,\text{Hz }\mu\text{G}^{-1}$ for \ion{H}{i}), this kind of study has only been carried out in a handful of regions hosting the strongest masers \citep[see the review by][]{Crutcher19}.
In general, magnetic field measurements in outflows, are extremely difficult to carry out, thus their role in the launching, collimation, and propagation of ionized jets has been scarcely studied from an observational point of view. 
In this sense, strong water masers are excellent tools to not only investigate the kinematics of the shocked gas, but also to get key information of the magnetic fields in the post-shock regions.

\medskip

In this manuscript, we present observations toward one of the sources in our sample, IRAS~19035$+$0641~A (I19035A hereafter), as a case study that demonstrates the potential of the full survey. 
I19035A is located at a distance of 2.2\,kpc \citep{Osterloh97} in a chemically rich environment \citep[e.g.,][]{Fuller05,Araya05,Araya07,Lopez-Sepulcre10,LopezSepulcre11,Lu14,Taniguchi19}. 
Based on HiGal data, \cite{Rosero19} report a bolometric luminosity of $6\times10^3\,$\lsun\ for this region. 
The ultra-compact (UC) \ion{H}{II} region IRAS\,19035+0641\,B \citep{Rosero16} is located at about 2\arcs\ to the South-East of I19035A.
The goal of this work is to investigate the jet nature of I19035A, hence we will exclude IRAS~19035+0641~B from the following analysis and discussion, but direct the reader to see Section 2.2 from \cite{Rosero19} for more information. 
This paper is organized as follows:
in Section~\ref{sec:observations}, the observations, data reduction, and imaging details are given. Our results are presented in Section~\ref{sec:results}, followed by a discussion in Section~\ref{sec:discussion}. Our findings are summarized in Section~\ref{sec:summary}, and complementary figures are given in the \hyperref[sec:appendix]{Appendix}.

\section{Observation \& Data Reduction}
\label{sec:observations}
The K-band ($18-26.5\,$GHz) observations were carried out with NRAO's\footnote{The National Radio Astronomy Observatory is a facility of the National Science Foundation operated under cooperative agreement by Associated Universities, Inc.} VLA on March 27, 2022 in the A configuration utilizing the 3-bit samplers. 
A total of 57 spectral windows (SPWs) with $128\times1\,$MHz channels were used for the continuum. A narrow 16\,MHz SPW with $1024\times15.6\,$kHz (i.e., $1024\times0.21\,$\kms) channels was set up to simultaneously observe the 22235.08\,MHz H$_2$O($6_{1,6}-5_{2,3}$) maser 
transition\footnote{From the Jet Propulsion Laboratory Molecular Spectroscopy Catalog \citep{JPL}, as referenced in the Splatalogue database (\url{https://www.cv.nrao.edu/php/splat/}).}. 
The target phase center was set to $\text{RA(J2000)}=19^{\text{h}}06^{\text{m}}01\overset{{\text{s}}}{.}1$, $\text{Dec(J2000)}=6\deg46^{\prime}35\overset{{\prime\prime}}{.}0$ and the total time on source was $\sim11$ minutes.
The primary beam and largest recoverable scale of these observations are $\sim2'$ and 2\farcs4, respectively.
Observations of the quasar 3C286 were used for flux density scale and bandpass calibration, while the complex gain calibration was done with observations of the calibrator source J1851+0035, with a cycle time of 4~minutes. 

The continuum data were processed through NRAO's Common Astronomy Software Applications (CASA, \citealt{casa}) Calibration Pipeline, version 6.2.1.7. The images were made using the CASA (version 6.5.2) task \texttt{tclean} a grid weighting approximately intermediate between natural and uniform (ROBUST = 0.5). The synthesized beam size, PA, and rms noise of the resulting continuum images are 0\farcs083$\,\times\,$0\farcs071, 1.75$^\circ$, and 13.5~\mujy, respectively. The central frequency of the final continuum image is 22.21\,GHz, i.e., 1.35\,cm (1.3\,cm continuum hereafter). 

The line SPW was calibrated using the CASA pipeline version 5.3.0-147, modified for spectral line observations. 
Self-calibration for both phase and amplitude was performed on the strongest water maser spot (maser spot \#4, see below) using NRAO's Astronomical Image Processing System (AIPS). 
The self-calibration solution tables were then applied to the entire line SPW data. 
We used the AIPS task \texttt{IMAGR} with a grid weighting intermediate between natural and uniform (ROBUST = 0) to create the Stokes \textit{I} water maser image cube for total intensity analysis. This resulted in a channel width, synthesized beam size, PA, and typical rms noise of 0.21 \kms, 0\farcs083$\,\times\,$0\farcs076, 8.1$^\circ$, and 13.2~\mjy per channel (in line-free channels), respectively. 
Finally, for the Zeeman effect analysis, we additionally made Hanning smoothed Stokes \textit{I} and \textit{V} cubes\footnote{$I=(RCP+LCP)/2$ and $V=(RCP-LCP)/2$, where RCP and LCP are the right and left circular polarization data, respectively.}.


\section{Results}
\label{sec:results}

\subsection{Radio Continuum Emission}
\label{sec:results-cont}
\begin{figure*}[!t]
    \centering
    \includegraphics[width=.48\textwidth]{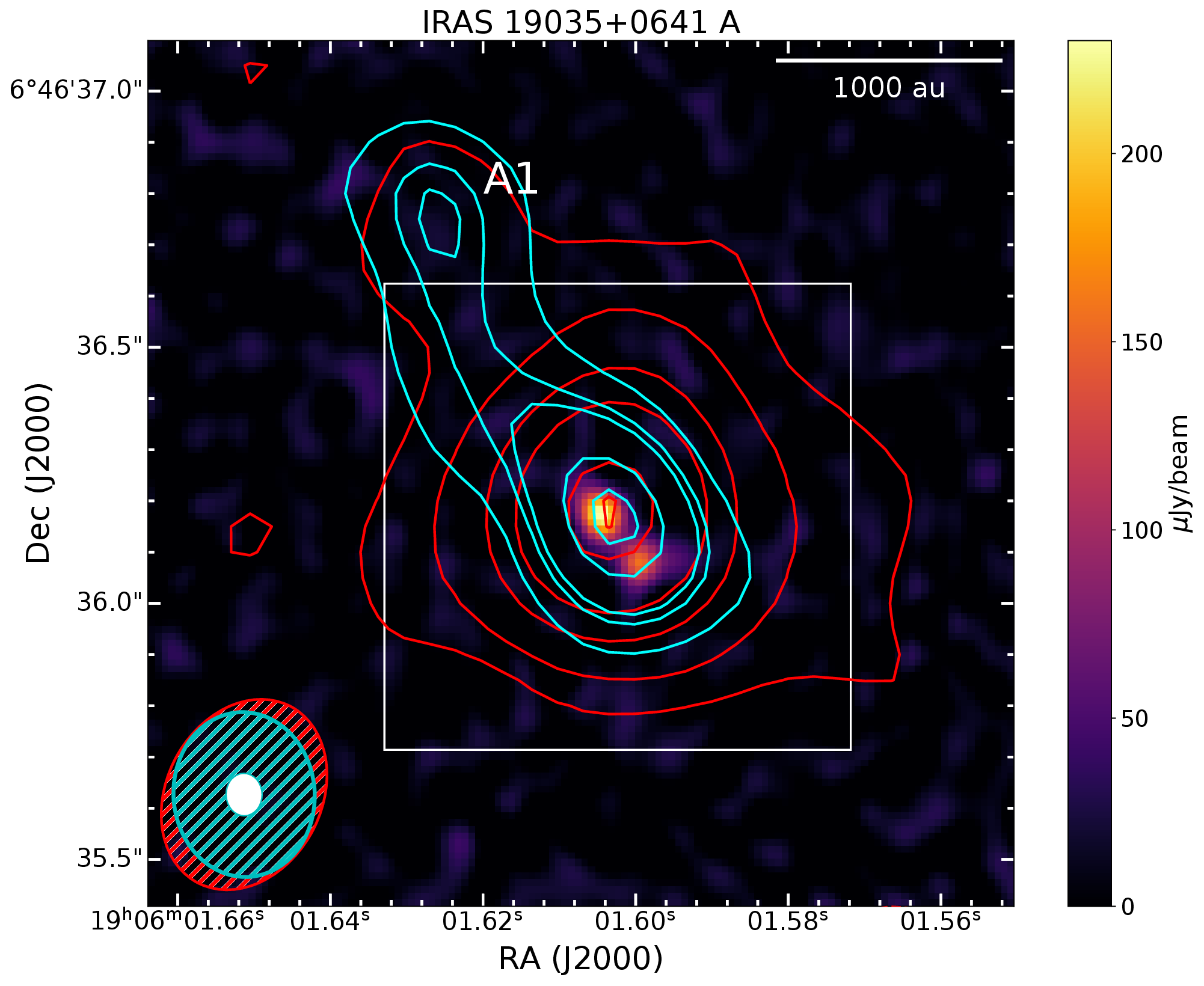}
    \includegraphics[width=.48\textwidth]{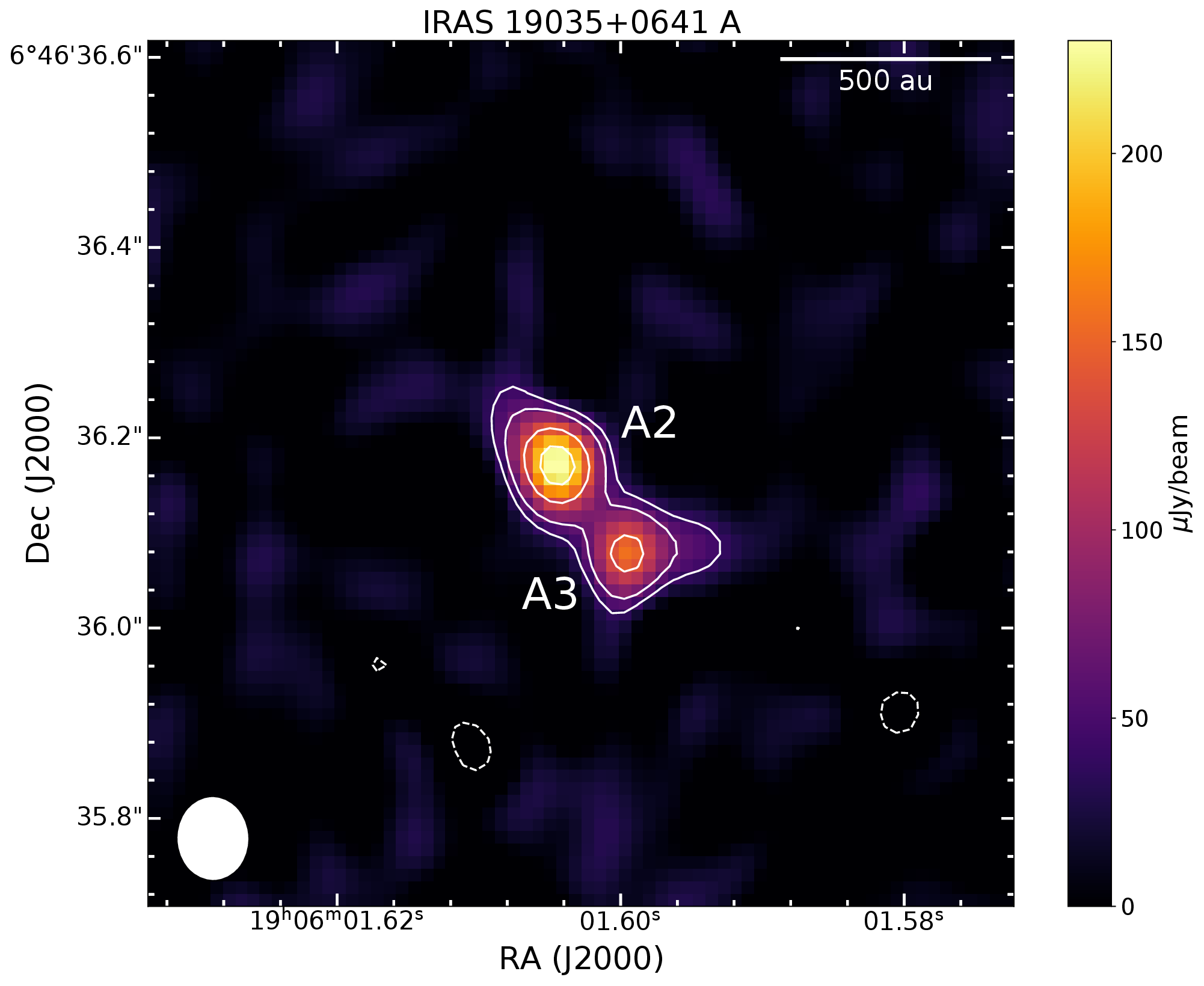}
    \caption{\textit{Left}: Our 1.3\,cm continuum emission toward I19035A is shown in color, and overlaid in red and cyan contours are the 1.3 and 6\,cm continuum emission from \cite{Rosero16}, respectively. Contour levels are $[-2,5,10,20,30,50,57]\times\,$7~\mujy\ and $[-3,3,5,10,12,20,25]\times4\,$\mujy\ for the K- and C-band data, respectively. 
    The ellipses in the bottom left represent our 1.3\,cm synthesized beam size (filled white, VLA A configuration), as well as the 1.3\,(red dashed, VLA B configuration) and 6\,cm (blue dashed, VLA A configuration) ones from \cite{Rosero16}. The white rectangle shows the region covered by the right panel.
    \textit{Right}: Our 1.3\,cm continuum emission toward I19035A in color and white contours. The latter show $[-3,3,5,10,15]\times13.5\,$\mujy\ levels.
    The filled white ellipse in the bottom right represents the synthesized beam size ($\sim0.08$\arcs).}
    \label{fig:continuum-overlay}
\end{figure*}

Our high angular resolution observations unveiled two 1.3\,cm continuum peaks, which are indicated as A2 and A3 in the right panel of Figure \ref{fig:continuum-overlay}, with a significance of $>5\sigma$ toward I19035A.
The positions of these peaks are listed in Table \ref{tab:cont-imfit}. Peaks A2 and A3 are aligned in the NE-SW direction, separated by $\sim$0\farcs12 ($\sim265$\,au), and were reported as a single radio continuum source at both 1.3 and 6\,cm by \cite{Rosero16} in their lower angular resolution observations. 
In addition, our detailed inspection of the images from the \cite{Rosero16} observations revealed a second, weaker 6\,cm continuum peak ($>10\sigma$) located $\sim0$\farcs65 ($\sim1400\,$au) to the NE of A2, as seen in the left panel of Figure~\ref{fig:continuum-overlay}.
We will refer to this second, weaker 6\,cm peak as A1. 
While the lower resolution 1.3\,cm continuum emission is dominated by a rather compact source, as reported by \cite{Rosero16}, it appears to be elongated toward A1 (see the lowest red contour in the left panel of Figure~\ref{fig:continuum-overlay}). However, this weak, extended 1.3\,cm continuum emission toward A1 was not recovered in our higher angular resolution observations.

The spectral index ($\alpha$, with $S_\nu\propto \nu^\alpha$) between the 6 and 1.3\,cm for the dominant continuum source, as seen in the lower resolution images, has a value of 0.9 \citep{Rosero16}. Through the re-inspection of their data, as noted above, we find that the spectral index flattens to 0 as we reach A1. In our high angular resolution 1.3\,cm continuum data, the measured in-band spectral indices are 0.9 and 0.6 for peaks A2 and A3, respectively.

We fitted 2-D Gaussians to each continuum peak using the highest resolution images available on them, that is, the 6\,cm for A1 and our 1.3\,cm for A2 and A3. The obtained positions, sizes, PA, peak intensities, and flux densities are presented in Table~\ref{tab:cont-imfit}. 
Peaks A2 and A3 are marginally resolved, with upper limits to their deconvolved minor axes sizes at full width half maximum (FWHM) as reported in the third column of the table. 
The sum of the flux densities of A2 and A3 is consistent with the value reported by \cite{Rosero16} for the 1.3\,cm continuum source in their image.

\begin{deluxetable*}{cccccccccc}[htpb]
\tablenum{1}
\label{tab:cont-imfit}
\tabletypesize{\scriptsize}
\tablecaption{Properties of the continuum peaks}
\tablecolumns{10}
\tablewidth{0pt}
\tablehead{
\colhead{Peak} & \colhead{RA$_{\mathrm{J2000}}$} & \colhead{Dec$_{\mathrm{J2000}}$} & \multicolumn2c{$\theta_{maj}$} & \multicolumn2c{$\theta_{min}$} & \colhead{PA} & \colhead{$I_\nu$} & \colhead{$S_\nu$}  \\
\colhead{} & \colhead{(h m s)} & \colhead{($^\circ\,'\,''$)} & \colhead{(mas)} & \colhead{(au)} & \colhead{(mas)} & \colhead{(au)} & \colhead{($^\circ$)} & \colhead{($\mu$Jy beam$^{-1}$)} &  \colhead{($\mu$Jy)}}
\startdata
A1 & 19 06 01.624 & 6 46 36.70 & 442 & 972 & 117 & 257 & 13.8 & 49 (5) & 91 (15) \\
\hline
A2 & 19 06 01.605 & 6 46 36.17 & 57 & 125 & $<21$ & $<46$ & 45.6 & 235 (13) & 277 (26)\\
A3 & 19 06 01.599 & 6 46 36.08 & 55 & 121 & $<11$ & $<24$ & 106.6 & 150 (13) & 165 (25) \\
\enddata
\tablenotetext{}{Values in parentheses are the uncertainties in the measurements. Note that the values for A1 were obtained using the C-band data from \cite{Rosero16}.}
\end{deluxetable*}

\subsection{22.235 GHz H$_2$O Masers}
\begin{figure*}
    \centering
    \includegraphics[width=.47\textwidth]{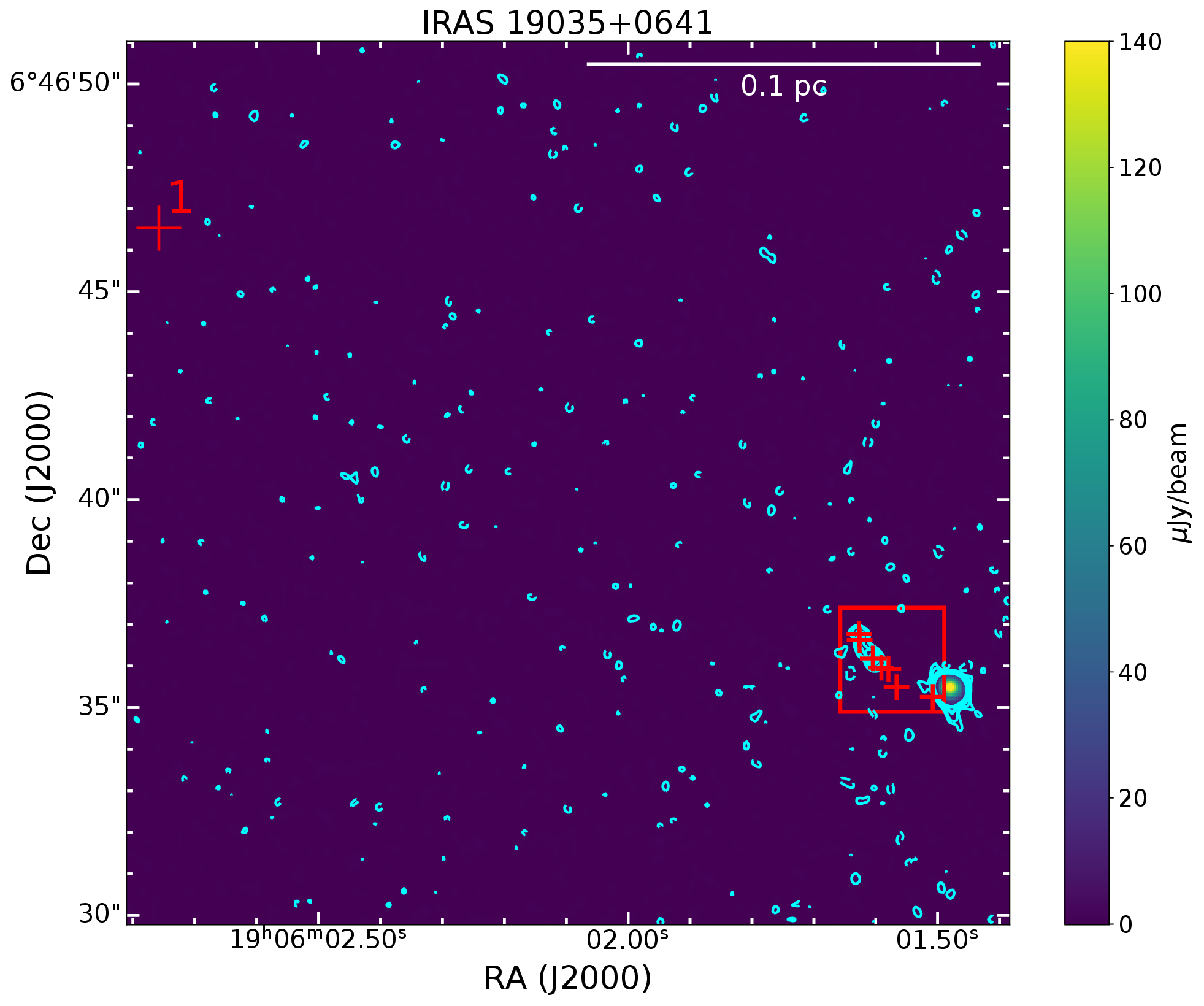}
    \includegraphics[width=.48\textwidth]{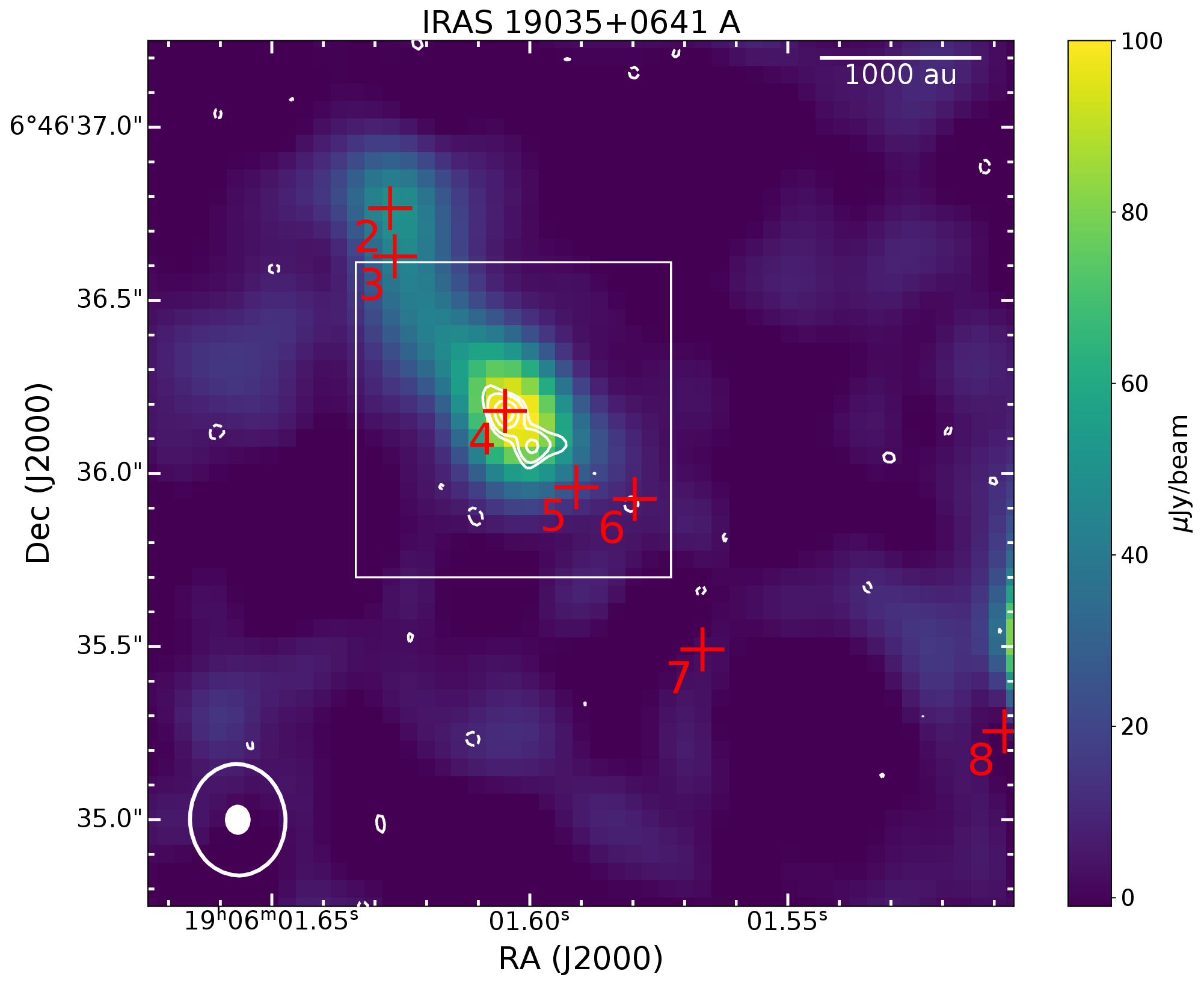}
    \caption{\textit{Left:} The 6\,cm continuum emission in IRAS~19035+0641 is shown in color and cyan contours \citep{Rosero16}. The contours are the same as in the left panel of Figure \ref{fig:continuum-overlay}.
    \textit{Right:} Enlarged version of the area marked with the red rectangle in the left-hand panel. The 6\,cm continuum emission is shown in color (note the scale difference between the panels), and our 1.3\,cm continuum emission is shown in white contours. The contours are the same as in the right panel of Figure \ref{fig:continuum-overlay}, and the outlined and filled ellipses in the bottom left represent the 6\,cm ($\sim$0\farcs33) and our 1.3\,cm ($\sim$0\farcs08) synthesized beam sizes, respectively. The white rectangle shows the region covered in the right panel of Figure \ref{fig:continuum-overlay}. In both panels, the red $+$ symbols mark the position of masers \#1 through \#8, which are also numbered.
    }
    \label{fig:continuum-masers}
\end{figure*}

We detected 8 water maser spots that seem to be associated with I19035A.
The maser spots are distributed along the same direction as the continuum emission and form a narrow linear structure in the NE-SW direction, as shown in Figure~\ref{fig:continuum-masers}. 
Maser \#1 is the farthest from the radio continuum source, located at about 0.2\,pc ($\sim$19\arcs) to the NE, while the other maser spots are found within $\sim$2500\,au from the continuum peaks. Masers \#2 and \#3 lay toward peak A1, while maser \#4 is found close to the center of A2. Masers \#5 $-$ \#8 are distributed to the SW of I19035A. 

We report the peak position and intensity of each maser spot in columns 2, 3, and 4 of Table \ref{tab:maser-pos}, respectively. We measured the velocity of the peak, its full width at zero power (FWZP), and the total velocity extent of the emission using the peak pixel spectrum for each maser spot. We list these in columns 5, 6, and 7 of Table \ref{tab:maser-pos}, respectively, while the spectra can be found in the \hyperref[sec:appendix]{Appendix}.
Maser spots \#4 and \#6 span a wide velocity range. This likely indicates that
these spots have substantial substructure that we are not able to spatially resolve.
The brightest maser spot, maser \#4, has a peak intensity of 102.8~Jy~beam$^{-1}$ at $V_{\mathrm{LSR}}^\mathrm{{peak}} = 20.1\,$\kms. 

\begin{deluxetable*}{ccccccc}[htpb]
\tablenum{2}
\label{tab:maser-pos}
\tabletypesize{\scriptsize}
\tablecaption{Properties of the Detected Maser Spots}
\tablecolumns{7}
\tablewidth{0pt}
\tablehead{
\colhead{\#} & \multicolumn2c{Peak Pos.}& \colhead{$I_\nu$} & \colhead{$V_{\mathrm{LSR}}^\mathrm{{peak}}$} & \colhead{$\Delta V_{\mathrm{FWZP}}^{\mathrm{peak}}$} & \colhead{$\Delta V$}  \\
\colhead{} & \colhead{RA (J2000)} & \colhead{Dec (J2000)} & \colhead{(Jy beam$^{-1}$)} & \colhead{(\kms)} & \colhead{(\kms)} & \colhead{(\kms)} 
}
\startdata
1&19h06m2.712s&6d46m45.98s&0.37 (0.01)&37.55& 1.26 &1.26\\
2&19h06m1.627s&6d46m36.77s&0.77 (0.01)&33.54& 2.95 &2.95\\
3&19h06m1.626s&6d46m36.62s&0.37 (0.01)&31.02& 1.05 &1.05\\
4&19h06m1.605s&6d46m36.18s&102.78 (0.19)&20.06&4.84 &35.60\\
5&19h06m1.591s&6d46m35.96s&0.29 (0.01)&37.12& 1.90 &1.90\\
6&19h06m1.579s&6d46m35.92s&3.61 (0.01)&36.07& 1.68&42.13\\
7&19h06m1.566s&6d46m35.49s&0.11 (0.01)&28.70& 1.05 &1.05\\
8&19h06m1.508s&6d46m35.25s&2.11 (0.01)&30.17& 1.90&3.37\\
\enddata
\tablenotetext{}{The columns show the maser spot number (\#), peak position, peak brightness ($I_\nu$), velocity of the peak brightness (V$_{\mathrm{LSR}}^{\mathrm{peak}}$), FWZP of the peak ($\Delta V_{\mathrm{FWZP}}^{\mathrm{peak}}$), and the total velocity extent of the maser spot emission ($\Delta V$). Values in parentheses are formal errors from the 2-D Gaussian fits.}
\end{deluxetable*}

\subsubsection{Zeeman splitting in maser \#4}
\begin{figure*}
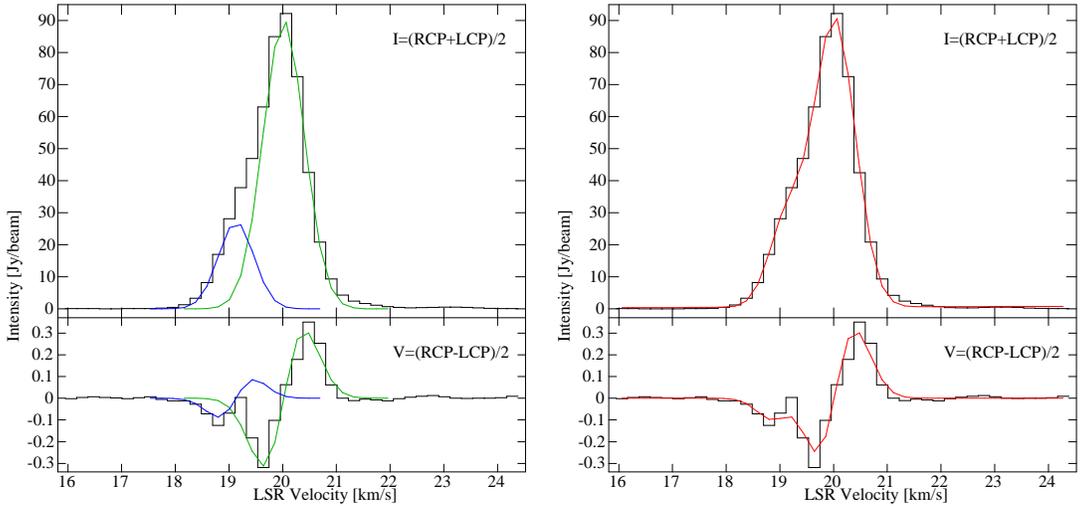

    \centering
    \includegraphics[width=.4\textwidth]{19035A_I_V_comp.eps}
    \includegraphics[width=.4\textwidth]{19035A_I_V_total.eps}
    \caption{Stokes \textit{I} (top) and \textit{V} (bottom) profiles of the brightest peak of maser spot~\#4 are shown with a black histogram-like line. The green and blue curves in the left panel show the two Gaussian components fitted to the Stokes \textit{I} profile (top) and their derivatives (bottom). The red curves in the right panel show the sum of the two components.
    }
    \label{fig:stokes-figures}
\end{figure*}

The Stokes \textit{I} and \textit{V} profiles of the strongest maser \#4 feature are shown with a histogram-like line in the top and bottom panels of Figure~\ref{fig:stokes-figures}, respectively.
The Stokes \textit{I} spectrum shows an asymmetric shape, indicative of at least two Gaussian components. 
We used the AIPS task \texttt{XGAUS} to fit the Stokes \textit{I} profile. The best fit, shown with a red line in the right panels of Figure~\ref{fig:stokes-figures}, corresponds to two Gaussian components, shown separately in the left panels with a blue and a green line. 
The S-shape appearance observed in the Stokes \textit{V} profile is usually considered a Zeeman effect detection. We derived the line-of-sight magnetic field $B_{\mathrm {los}}$ by fitting the Stokes \textit{V} profile \citep[e.g.,][]{Momjian17,Momjian19} to the expression \citep{Troland82,Sault90}:
\begin{equation}
\label{eq:stokesV}
    V=aI+\frac{b}{2}\frac{dI}{d\nu}.
\end{equation}
The first term on the right-hand side of the equation accounts for small leakage terms and, in this case, the scale factor $a$ was on the order of $10^{-3}$. The second term is the frequency derivative of the Stokes \textit{I} profile multiplied by a fit parameter $b$, which contains the magnetic field information. Namely, $b=zB_{\mathrm {los}}$, where \textit{z} is the Zeeman splitting factor. For the 22.2\,GHz water transition, $z=2.1\,\text{Hz mG}^{-1}$ \citep{Nedoluha92}.
We used the AIPS task \texttt{ZEMAN} to fit equation~\ref{eq:stokesV}, which derives a \textit{b} value for each Gaussian component by fitting their derivative to the Stokes \textit{V} profile.
As seen in Figure~\ref{fig:stokes-figures}, the Stokes \textit{V} profile of maser \#4 is well represented by the derivatives of the fitted Gaussians, and the measured $B_{\mathrm {los}}$ values are $123\,(\pm 27)$ and $156 \,(\pm 8)$~mG. 
These values are at a level of $4\sigma$ and $19\sigma$ (respectively), hence the measured magnetic field is significant in both cases.
We note that, by convention, a positive value for $B_{\mathrm {los}}$ means the line-of-sight magnetic field is pointing away from the observer.

\subsubsection{Follow-up observations: extreme H$_2$O maser variability}
\label{sec:follow-up}
The detection of the $\sim100$ Jy H$_2$O maser peak in I19035A led us to investigate the Zeeman splitting of this line and the magnetic field in the post-shock region. 
However, since the original observations were not designed for Zeeman effect analysis (e.g., the use of 3-bit samplers), we carried out follow-up VLA B-configuration observations on March 14, 2023, to confirm the Zeeman splitting. These observations, which utilized the 8-bit samplers, resulted in $\sim$ 30\,minutes of on-source time. 
We used a single 8\,MHz wide SPW with $2048\times3.91\,$kHz channels, which translates to a $52\,\text{m s}^{-1}$ channel width.
The data were manually calibrated \replaced{using CASA and the}{and imaged using CASA. The} synthesized beam size and typical rms noise of the final image cube are $\sim$0\farcs3 and $5.5\,$\mjy\ per channel in line-free channels, respectively. 

\begin{figure*}
    \centering
    \includegraphics[width=.8\textwidth]{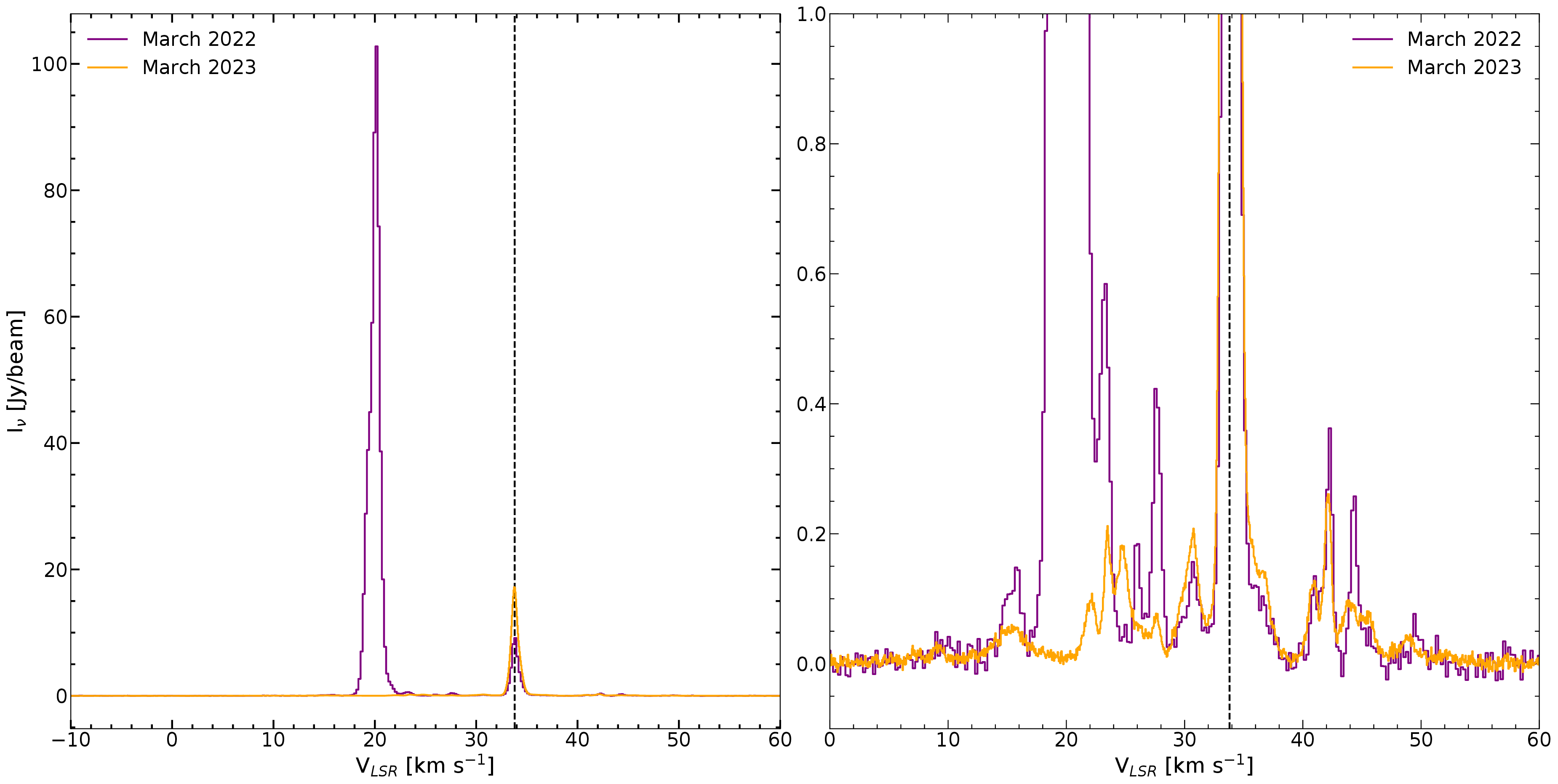}
    \caption{Comparison of maser \#4 peak spectrum as observed in March 2022 (purple) and March 2023 (orange). The left and right panels show the full and zoomed in spectra, respectively. All the spectral features show different degrees of variability. Particularly, the strong $\sim100\,$Jy peak was not detected in the March 2023 observations. The vertical dashed line marks the adopted systemic velocity \citep[$V_{sys}=33.8$ \kms,][]{Araya05}.
    }
    \label{fig:variability}
\end{figure*}

The new observations recovered most maser spots seen in the 2022 observations, with varying degrees of variability. In Figure~\ref{fig:variability}, we show a comparison of the two epochs for maser spot \#4. While most peaks observed in the original spectrum are present, albeit with different intensities,
the bright feature that showed Zeeman splitting was not detected in the new data. 
Thus, while we were not able to confirm the Zeeman splitting detection, the complete disappearance of the $\sim$100\,Jy peak in maser \#4 indicates extreme H$_2$O maser variability in I19035A.

\section{Discussion}
\label{sec:discussion}

\subsection{Nature of the Radio Continuum Emission}
\label{sec:discussion-jet}
Based on the compact nature of our new 1.3\,cm continuum data, one might consider the possibility of peaks A2 and A3 being two individual YSOs in a tight ($d\approx265\,$au) protobinary system \citep[e.g.,][]{Kraus17,Zhang19}. 
In this scenario, the radio continuum emission would be attributed to early-stage \ion{H}{II} regions.
If A2 and A3 have a spherical geometry of diameter equal to their geometric mean (calculated using the sizes reported in Table \ref{tab:cont-imfit}), then their radii are approximately 40 and 30\,au, or about $1.5\times10^{-4}$ pc. 
These sizes are consistent with those expected for HC \ion{H}{II} regions \citep[$<0.03\,$pc,][]{Kurtz05}. 
Using the ionizing photon rate ($N_{Ly}$) equation from \cite{Sanchez-Monge13}, assuming optically thin emission from a homogeneous \ion{H}{II} region and an electron temperature of $10^4$ K, we calculate $N_{Ly}$ values of $1.5\times10^{44}$ and $9.5\times10^{43}\,\text{s}^{-1}$ for A2 and A3, respectively, corresponding to B2-B3 Zero Age Main Sequence stars. 

Note that the cm spectrum is still rising in K-band, which for the above calculation was assumed to mark the turnover point between optically thick to optically thin conditions.
A rising spectrum can also be explained by a stratified electron density distribution, which naturally would occur in an ionized stellar wind.
We can use equation 5 in \cite{Felli82} to estimate the stellar wind mass loss rate. Assuming a terminal wind velocity of 600 \kms,
we obtain mass loss rates of the order of $10^{-7}$ \msun\ yr$^{-1}$. This number is several magnitudes higher than what is expected from near-main-sequence B2-B3 type stars \citep[e.g.,][]{Cohen97}. Nonetheless, for stars in evolutionary stages prior to the main sequence, such high values are perhaps not unexpected.

While the above analysis shows that our data are consistent with a model where the radio continuum peaks A2 and A3 are caused by a high-mass binary, there is much morphological
evidence that an ionized jet model is preferable.
First, the observed extended and elongated morphology of the radio continuum emission at 6 cm can hardly be attributed to two compact objects located at the position of A2 and A3. Moreover, this elongated structure is very well aligned with A2 and A3, in the NE-SW direction.
Second, the water maser spots \#2 $-$ \#7 are also very well aligned in the direction of the radio continuum emission, along a structure much larger than the distance between A2 and A3, which is inconsistent with a wind-driven shell scenario \citep[e.g.,][]{Beltran07}. 
Ultimately, even though we cannot entirely discard a high mass binary for the origin of the observed 1.3 cm continuum emission, it appears more likely to be tracing an ionized jet. We will now discuss this scenario.

\cite{Rosero19} had classified I19035A as a possible ionized jet based on its association with a bipolar molecular outflow observed in CO(2$-$1) and HCO$^+(1-0$) \citep{Beuther02b,Lopez-Sepulcre10}. The orientation of the outflow at the angular resolution of these studies (11\arcs\ and 29\arcs, respectively) is SE-NW, i.e., perpendicular to the NE-SW orientation of the 6\,cm continuum.
However, the classification as a high-mass ionized jet was still made based on a number of additional observational results, like the coincidence of the $6\,$cm continuum with a cone-shaped 2.2$\,\mu$m structure, presumably tracing scattered light from an outflow cavity, the presence of SiO(2$-$1) and (3$-$2) in the region \citep{LopezSepulcre11}, as well as detections of Class II 6.7\,GHz CH$_3$OH masers in the vicinity of I19035A \citep[]{Beuther02c,Hu16,Ouyang19}. 
Furthermore, with a radio luminosity of $S_{5GHz} d^2 =1.05\,$mJy~kpc$^2$, I19035A falls in the high-mass end of the $S_{\nu} d^2 - L_{Bol}$ relation shown in figure 8 from \cite{Rosero19} for ionized jets, and it is well below the radio luminosity values for UC/HC \ion{H}{ii} regions ($\approx10^2\,$mJy~kpc$^2$). This further indicates that the I19035A radio luminosity at 5\,GHz is likely due to shock ionization rather than photoionization from the central YSO. 
The fact that the large scale outflow is perpendicular to the jet direction can be explained by precession of the jet axis, for which there are a number of known cases discussed in the literature \citep[e.g.,][]{Rodriguez21}. 
High angular resolution observations of the bipolar molecular flow are necessary to verify this conjecture and unveil the true outflow orientation.

Recently, \cite{Fedriani23}, using the Stratospheric Observatory for Infrared Astronomy (SOFIA) $7.7-37.1\,\mu$m observations together with multi-wavelength data from the literature, compiled the detailed SED of IRAS\,19035+0641 and carried out radiative transfer modeling based on the turbulent core accretion theory \citep[][and references therein]{Zhang18}. Their average model predicts a central protostellar mass of $15\,$M$_\odot$ accreting at a rate of
$2.1\times 10^{-4}\,$M$_\odot\,$yr$^{-1}$, and a bolometric luminosity of $4.9\times10^4\,$\lsun. The mass outflow rate 
of the CO was estimated by \cite{Beuther02b} as $0.9\times 10^{-4}\,$M$_\odot\,$yr$^{-1}$, i.e., consistent with the idea that part of the energy released by the accretion process powers the outflow.

What do our new continuum data add to this picture? 
The radio luminosity at 5 GHz discussed above was derived from a source consisting of a compact core with a NE-SW extension \citep{Rosero16}.
The higher angular resolution $1.3\,$cm observations reported here resolve the source into two peaks (A2 and A3) aligned in the NE-SW direction. This shows that the elongated bipolar structure persists on a much smaller scale of $\lesssim 500\,$au, as expected from an ionized jet. Moreover, the rising in-band spectral indices for both A2 and A3 (0.9 and 0.6, respectively) are in good agreement with the ideal jet model from \cite{Reynolds86}. 
In conclusion, our observations strengthen the interpretation that I19035A is an ionized jet from a high-mass protostar.

\subsection{The Water Masers}
\subsubsection{Association with the I19035A Jet}
\begin{figure}
    \centering
    \includegraphics[width=.45\textwidth]{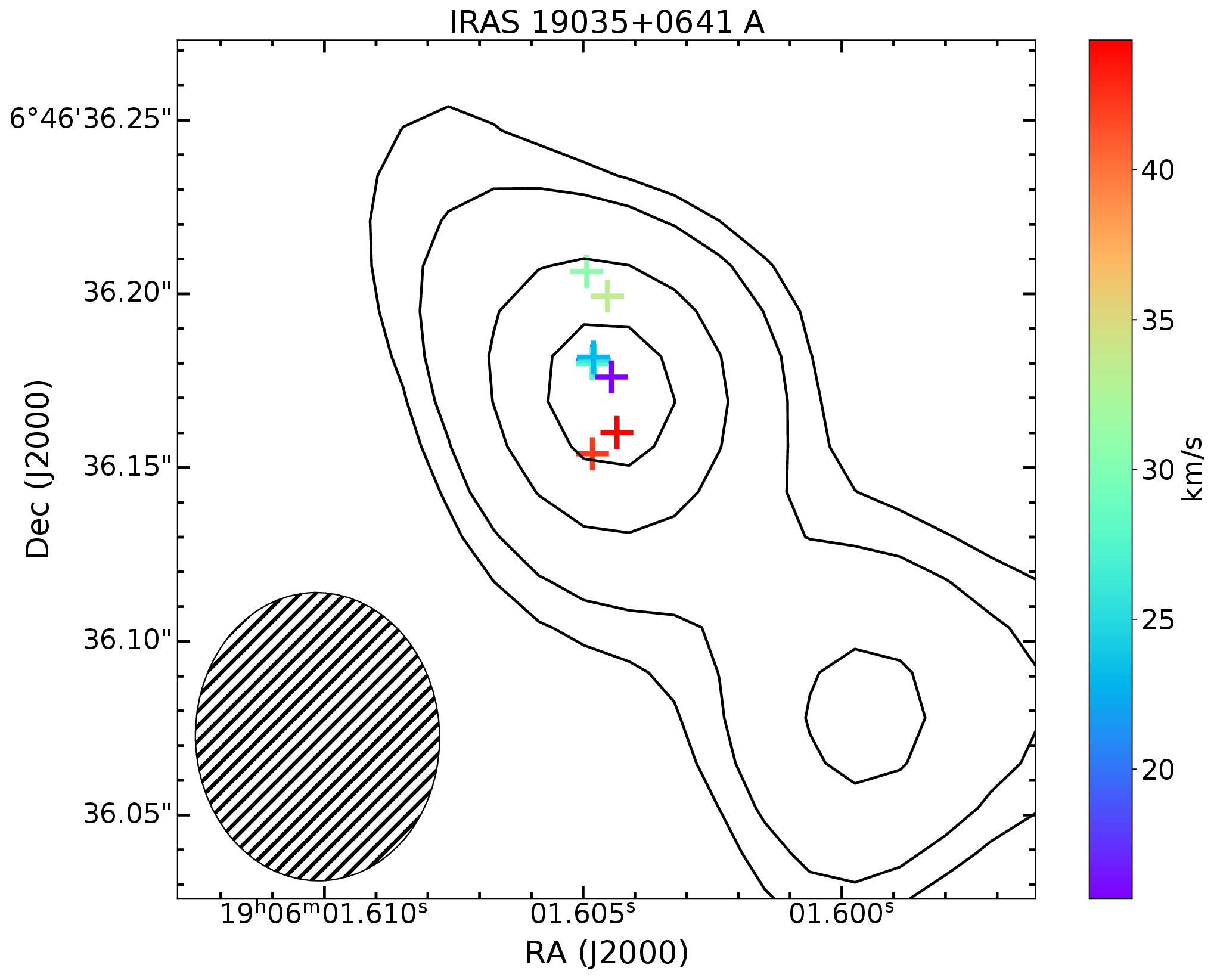}
    \caption{Positions obtained from 2-D Gaussian fits to each maser \#4 spectral feature are shown with $+$ symbols overlaid to the 1.3~cm continuum in black contours. The data points are color-coded by velocity the contours are the same as in the right panel of Figure~\ref{fig:continuum-overlay}. The dashed black ellipse in the bottom left represents the synthesized beam size.
    We note that the difference in the positions obtained are much larger than the formal fit errors provided by the CASA task \texttt{imfit}.
    }
    \label{fig:maser4-scatterplot}
\end{figure}

Water masers in high-mass star forming regions can be associated with both jets/winds and accretion disks \citep[e.g.,][]{Imai06,Sarma08,Moscadelli22}.
It is therefore pertinent to analyze whether the water masers detected in I19035A probe the ionized jet or a disk.
In the latter case, water maser spots arise in different regions of the disk, and sometimes show Keplerian velocity gradients and arc-shape structures of a few au in size that are only resolved using Very Large Baseline Interferometry (VLBI) techniques \citep[e.g.,][]{Torrelles98,Franco-Hernandez09,Rodriguez12}. 
In contrast, our observations reveal a well defined linear maser spot distribution covering about 2500 au (excluding maser \#1, located at $\sim40\,000$ au from the continuum source), and the water maser positions are well aligned with the ionized jet axis. If these were associated with a disk, a distribution perpendicular to the jet would be expected. 
Therefore, the water masers in I19035A appear to be mostly associated with the ionized jet observed in the radio continuum. 
However, we note that maser spot \#4, located close to the radio continuum peak A2, shows several emission peaks of varied intensity at red- and blue-shifted velocities from the adopted $V_{LSR}$ (see Figure~\ref{fig:variability}), thus it deserves additional consideration. 
We carried out 2-D Gaussian fits to each maser \#4 spectral feature to obtain their positions, which we show color-coded by velocity on top of the 1.3\,cm continuum emission in Figure \ref{fig:maser4-scatterplot}. 
The individual components of maser spot \#4 are located along a N-S line, and there is an apparent velocity gradient along this structure. 
If this gradient was associated with a Keplerian accretion disk, the equation $M=RV^2/G$ can be used to estimate the mass of the central object. Using this expression, the derived mass is $\sim95\,$\msun, which appears  too large for an early stage high-mass YSO. Therefore, we believe it is unlikely that maser spot \#4 traces an accretion disk. More likely, the observed velocity gradient is reflecting the velocities of individual shocks in the outflowing gas, and might be tracing an expanding or rotating flow. VLBI observations and a proper motion study are required to determine the nature of maser spot \#4.

\medskip 

In the following, we discuss the association of the maser spots \#1 and \#8 with the I19035A jet. 
Maser \#1 is found at $\sim0.2\,$pc ($\sim40000\,$au) from the ionized jet, thus their relation is not immediate. 
Although \cite{Moscadelli20} found that 84\% of water masers are found within 1000\,au from the driving protostar, they report detections at distances up to $\lesssim18000\,$au (0.09\,pc), and strong jet bow-shocks where water masers can be pumped are usually found at pc-scale distances from the high-mass YSOs \citep[e.g., HH\,80/81 at $\sim$1\,pc from IRAS~18162$-$2048,][]{Reipurth88}. 
As seen in Figure~\ref{fig:continuum-masers}, the PA of maser \#1 is consistent with that of the ionized jet and the other water maser spots.  
Also the peak velocity of maser \#1 is within 10~\kms\ from the systemic velocity of the region, and it is consistent with the velocity range in which the other masers in the jet present emission. Moreover, we detected variability in the intensity of maser \#1 in our follow up observations (see Section \ref{sec:follow-up}), which suggests a common origin for this and the other maser spots in the jet. 
Finally, we did not detect a radio continuum source in the vicinity of maser \#1 that could be associated with its emission, although we do not discard the possibility of this source being outside our field of view or fainter than $\sim40\,$\mujy\ (i.e., our $3\sigma$ limit).
Meanwhile, maser \#8 is found in close proximity (projected) to the UC \ion{H}{ii} region IRAS~19035+0641~B.
The association of water masers with \ion{H}{ii} regions has been investigated in the past. Observational studies found that water masers can be actively pumped in the edges of expanding wind-driven, ionized shells \citep[e.g.,][]{Beltran07}. 
It has also been suggested that they might coexist with \ion{H}{ii} regions and eventually disappear while the YSO remains visible \citep[e.g.,][]{Sanchez-Monge13}. 
The PA and peak velocity of maser \#8, as in the case of maser \#1, are consistent with the other water masers in the jet. We did not detect maser \#8 in our follow-up observations, which might be indicating extreme variability and a common pumping mechanism with the other masers that arise in the jet. Additionally, maser spot \#8 was the only one detected nearby the UC\,\ion{H}{II} region, and we do not see an arc- or ring-like distribution, indicative of an expanding shell, around IRAS~19035+0641~B. This could be due to an anisotropic shell structure, however, there is not enough information in the literature about the kinematics of this UC\,\ion{H}{II} region, such as the expanding shell velocity.
In summary, our observations suggest maser \#1 is likely associated with the I19035A jet and are insufficient to unambiguously determine the origin of maser \#8. A subsequent water maser proper motion study could shed light on the connection between these masers and the jet. Alternatively, observations of other shock probes and molecular outflow tracers could be used to determine whether there is a spatial correlation between masers \#1 and \#8 and the jet/outflow structure.

\subsubsection{Estimation of Energetic Parameters from the Measured Magnetic Field}
\label{sec:discussion-energetics}
We can use the more significant line-of-sight magnetic field component of $B_{\mathrm {los}} \sim 156$\,mG, measured from our possible Zeeman splitting detection to estimate some interesting jet and shock parameters. 
Firstly, we will estimate the number density ($n_{\mathrm {post}},n_{\mathrm {pre}}$) and magnetic field strength ($B_{\mathrm {post}},B_{\mathrm {pre}}$) in the pre- and post-shock gas. 
If we assume that the magnetic field is amplified proportionally with the density in the shocked region \citep[e.g.,][]{Sarma02}, then we can relate these values as:
\begin{equation}
    \label{eq:Bpre}
    \frac{B_{\mathrm {post}}}{B_{\mathrm {pre}}}=\frac{n_{\mathrm {post}}}{n_{\mathrm {pre}}}.
\end{equation}
In addition, \cite{Crutcher99} showed that, statistically, the total post-shock magnetic field ($B_{\mathrm {post}}$) can be well approximated by $B_{\mathrm {post}}=2B_{\mathrm {los}}$. 
We can also relate the $B_{\mathrm {pre}}$ and the pre-shock gas density ($\rho_{\mathrm {pre}}$) as:
\begin{equation}
    \label{eq:B0-rho0}
    B_{\mathrm {pre}}=\beta\rho_{\mathrm {pre}}^{1/2},
\end{equation}
where $\beta\approx 5.8\times10^5$ in cgs units \citep[calculated empirically by][]{Sarma02}. 
Combining these expressions, and taking $n_{\mathrm {post}}\sim10^9\,\text{cm}^{-3}$ \citep{Elitzur92}, we find that the density in the pre-shock region is $\rho_{\mathrm {pre}}=n_{\mathrm {pre}}m\sim10^{-16}\,$g~cm$^{-3}$, where $m$ is the mean molecular weight (i.e., $m=2.8m_p$, with $m_p$ being the proton mass). 
This yields a pre-shock gas number density of the order of $n_{\mathrm {pre}}\sim2\times10^7\,\text{cm}^{-3}$, which in turn translates to a pre-shock magnetic field strength of $B_{\mathrm {pre}}\sim7\,$mG.

We can also investigate what process is more likely to dominate the motion in the post-shock region by comparing the magnetic energy density ($\mathcal{M}$) and the kinetic energy density ($\mathcal{K}$).
The magnetic energy density is defined as:
\begin{equation}
    \mathcal{M}=B_{\mathrm {post}}^2/8\pi.
\end{equation}
\cite{Crutcher99} showed that, for an ensemble of measurements, a good approximation for the post-shock magnetic field value squared is given by $B_{\mathrm {post}}^2=3B_{\mathrm {los}}^2$. 
For $B_{\mathrm {los}}=156$\,mG, $\mathcal{M}=2.9\times10^{-3}\,\text{erg cm}^{-3}$.
Meanwhile, the kinetic energy density can be expressed as:
\begin{equation}
    \mathcal{K}=(3/2)mn_{\mathrm {post}}\sigma^2,
\end{equation}
where $\sigma$ is the velocity dispersion of the FWHM linewidth $\Delta v$, i.e., $\sigma=\Delta v(8\ln2)^{-1/2}$. 
If $\Delta v$ is the FWHM of the brightest peak in maser \#4, which is $1.16\,$\kms, then $\mathcal{K}\sim 1.7\times10^{-5}\,\text{erg cm}^{-3}$ in the post-shock region. 
However, the maser emission is amplified along the path of greatest velocity coherence, which translates to a narrowing of the maser line.
Therefore, a $\Delta v$ value obtained from a maser line is a lower limit of the real dispersion of the gas in the region. 
If instead we use $\Delta v=7.0\,$\kms\ from single-dish CH$_3$CN observations \citep{Araya05}, 
the kinetic energy density would be of the order $\mathcal{K}\sim6.2\times10^{-4}\,\text{erg cm}^{-3}$. This is still an order of magnitude smaller than $\mathcal{M}$, indicating that the magnetic energy is likely dominating the motion in the post-shock region.

In addition, the shock velocity $v_s$ can be estimated by equating the magnetic energy density and the ram pressure \citep[e.g.,][]{Sarma08}:
\begin{equation}
    B_{\mathrm {post}}^2/8\pi=\rho_{\mathrm {pre}} v_s^2.
\label{eq:v_shock}
\end{equation}
For $B_{\mathrm {los}}=156\,$mG and the $\rho_{\mathrm {pre}}$ estimated above, we obtain a shock velocity of 62~\kms.
Typically, shocks are classified according to their velocity as \textit{C}- or \textit{J}-type shocks, with the former having $v_s\lesssim30\,$\kms\ and the latter $v_s>30\,$\kms. 
The derived shock velocity values are in good agreement with maser spot \# 4 arising in a \textit{J}-type shock.

\bigskip

As noted previously, the above calculations make use of the $B_{\mathrm {los}}$ measurement obtained from our Zeeman splitting detection in the brightest peak from the maser \#4 spectrum. 
Even though we were unable to confirm this detection, as noted in Section~\ref{sec:follow-up}, we briefly discuss the implications of the calculated physical parameters on the jet. 
The estimated magnetic field in the jet is $B_{\mathrm{pre}}\sim 7\,$mG, which is comparable to the higher values obtained by \cite{Curran07} for a set of high-mass YSOs at scales of about 0.1\,pc ($\mathbf{<0.1\sim6}\,$mG). 
We expect the value of the magnetic field along the jet axis to remain fairly constant at scales of $\sim0.5\,$pc \citep[e.g.,][]{Carrasco10}, therefore it is likely that the energetic parameters derived above represent the magnetic field in the I19035A jet.
However, when comparing the magnetic energy density and the kinetic energy density derived from the pre-shock magnetic field ($B_{\mathrm {pre}}\sim7\,$mG) and density ($n_{\mathrm {pre}}\sim2\times10^7\,\text{cm}^{-3}$) values, we find that the magnetic energy density ($\mathcal{M_{\mathrm{pre}}}\sim 1.9\times10^{-6}\,\text{erg cm}^{-3}$) is an order of magnitude smaller than the kinetic energy density in the pre-shock region ($\mathcal{K_{\mathrm{pre}}}\sim 1.2\times10^{-5}\,\text{erg cm}^{-3}$).
Therefore, while the magnetic field appears to dominate the post-shock gas dynamics, the motions along the jet seem to be dominated by the gas kinematics.

\subsubsection{Maser Variability}
There has been increasing observational evidence that variations in the underlying protostellar object could be responsible for maser flares and maser variability of several different species.
\cite{Brogan18} discuss the case for water maser variability in an outflow caused by a burst of infrared radiation due to an accretion event propagating along the flow cavity. In this scenario, extreme increase or decrease in water maser intensity has a radiative origin, even though the principal pumping mechanism is collisions (see also \citealt{Gray16}). 
A similar scenario was invoked by \cite{Hirota21} to explain water maser variability in the high-mass protostellar object S255\,NIRS\,3. 
Alternatively, \cite{Trinidad03} demonstrate that random Gaussian fluctuations of an unsaturated maser (i.e., exponential amplification) in the line opacity can account for maser variability in the span of a few months. The results in \cite{Andreev17} are also consistent with changes in the amplification path length.

In I19035A, we detected strong variability within one year in all water maser spots, although we do not see a global increase or decrease trend in the intensities. In all cases, some spectral features become brighter and others dimmer, new lines appear, and others become completely undetected. No position dependence of the variability along the NE-SW line of H$_2$O masers was found either. 
While we find many very small scale intensity variations, 
the weakening of a $\sim 100\,$Jy maser to below $16.5\,$mJy ($3\times$ the rms noise in our follow-up observations) makes a statistical fluctuation unlikely.
A possible scenario that would explain the observed variations is that accretion events occurred in I19035A. Monitoring of the H$_2$O maser line, VLBI observations, and observations of other probes (such as the radiatively pumped Class II CH$_3$OH masers) toward I19035A are needed to explore in detail the origin of the water maser variability in this high-mass protostellar object.

\section{Summary \& Conclusions}
\label{sec:summary}
We report VLA A-configuration observations of the 1.3\,cm continuum and 22.2\,GHz H$_2$O masers toward the high-mass protostar I19035A. 
Our discoveries are summarized as follows:

\begin{enumerate}
    \item We resolved the 1.3~cm continuum emission toward I19035A into a bipolar structure consisting of two peaks aligned in the NE-SW direction. We showed that the elongated, bipolar structure reported by \cite{Rosero16,Rosero19} persists at scales $\lesssim500\,$au. The in-band spectral index calculated is consistent with what is expected from an ionized jet. 
         
    \item We detected eight 22.2~GHz H$_2$O maser spots well aligned along the ionized jet axis. The intensity of the brightest peak in each maser spot ranges between 0.11 (maser \#7) and 102.78~Jy~beam$^{-1}$ (maser \#4).  

    \item We detected possible Zeeman splitting in the brightest line observed. 
    The Stokes \textit{I} profile was well fitted by two Gaussian components, and the measured line-of-sight magnetic field values are 123 and 156\,mG. This yields a pre-shock magnetic field of $\sim7\,$mG.
    
    \item We carried out a second set of observations of the water maser emission in I19035A spaced one year from the original to confirm our Zeeman splitting detection. We found intense variability in all maser spots. Particularly, the bright $\sim100\,$Jy~beam$^{-1}$ peak was not detected. 
    
\end{enumerate}

In conclusion, our observations of I19035A strengthen the original classification as an ionized jet from a high mass protostar based on the bipolar nature of the 
$1.3\,$cm continuum on scales $\lesssim 500\,$au, the occurrence of H$_2$O masers along the jet axis, and the presence of extreme variability which could have been caused by an accretion event.
\\
\\
\small{
\textit{Acknowledgments.} We thank the anonymous reviewer for their excellent comments and suggestions that improved this work. T.M.R. is supported by the NRAO's Grote Reber Doctoral Fellowship. P.H. and E.D.A. acknowledge support from NSF grants AST–1814011, and AST–1814063, respectively. This work was partially funded by the NRAO's Student Observer Support (T.M.R. and P.H.).
}

\bibliography{main}{}
\bibliographystyle{aasjournal}

\begin{appendix}
\label{sec:appendix}

\setcounter{figure}{0}
\renewcommand\thefigure{A.\arabic{figure}}

\begin{figure*}[b]
    \centering
    \includegraphics[width=0.98\textwidth]{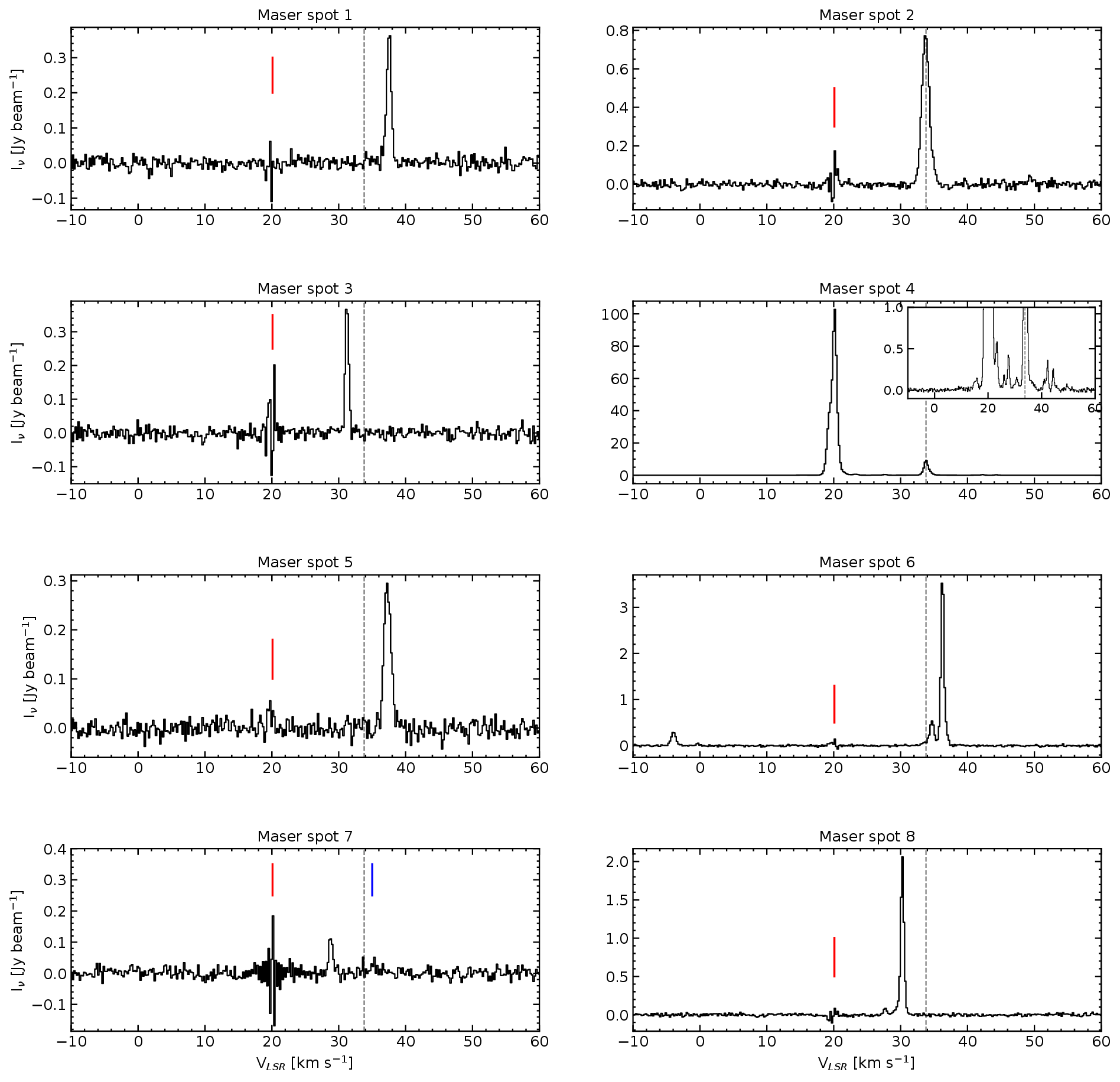}
    \caption{Peak pixel spectra of all the detected maser spots. The vertical dashed line marks the adopted systemic velocity \citep[$V_{sys}=33.8$ \kms,][]{Araya05}. The solid red and blue lines represent residual side lobes contamination from masers \#4 and \#6, respectively.
    }
    \label{fig:app-spectra}
\end{figure*}

\end{appendix}
\end{document}